\documentclass[a4paper,fleqn,usenatbib]{mnras}

\usepackage[T1]{fontenc}
\usepackage{ae,aecompl}

\usepackage{graphicx}	
\usepackage{amsmath}	
\usepackage{amssymb}	
\usepackage{float}
\usepackage{caption}

\begin{document}
\label{firstpage}
\pagerange{\pageref{firstpage}--\pageref{lastpage}}

\title[The Discovery of Four Massive DAVs]{Four New Massive Pulsating White Dwarfs Including an Ultramassive DAV}

\author[Brandon Curd et al.]{
Brandon Curd$^{1}$, 
A. Gianninas$^{1}$, 
Keaton J. Bell$^{3,4}$, 
Mukremin Kilic$^{1}$, 
\newauthor 
A. D. Romero$^{5}$, 
Carlos Allende Prieto$^{6,7}$,
D. E. Winget$^{3}$,
and
K. I. Winget$^{3}$
\\
$^{1}$Homer L. Dodge Department of Physics and Astronomy, University of Oklahoma, Norman, OK 73019, USA\\
$^{2}$Harvard-Smithsonian Center for Astrophysics, 60 Garden St., Cambridge, MA 02138, USA\\
$^{3}$Department of Astronomy, University of Texas at Austin, Austin, TX 78712, USA\\
$^{4}$McDonald Observatory, Fort Davis, TX 79734, USA\\
$^{5}$Departamento de Astronomia, Universidade Federal do Rio Grande do Sul, Av.  Bento Goncalves 9500, Porto Alegre 91501-970, RS, Brazil\\
$^{6}$Instituto de Astrof\'{\i}sica de Canarias, E-38205, La Laguna, Tenerife, Spain\\
$^{7}$Departamento de Astrof\'{\i}sica, Universidad de La Laguna, E-38206, La Laguna, Tenerife, Spain
}

\date{Accepted XXX. Received YYY; in original form ZZZ}

\pubyear{2016}

\maketitle

\begin{abstract}
We report the discovery of four massive ($M > 0.8\,M_\odot$) ZZ Ceti white dwarfs, including an ultramassive $1.16\,M_\odot$ star.
We obtained ground based, time-series photometry for thirteen white dwarfs from the Sloan Digital Sky Survey Data Release 7 and Data Release 10 whose
atmospheric parameters place them within the ZZ Ceti instability strip.
We detect mono-periodic pulsations in three of our targets (J1053, J1554, and J2038) and 
identify three periods of pulsation in J0840 (173, 327, and 797 s). 
Fourier analysis of the remaining nine objects do not indicate variability above the $4\langle{A}\rangle$ detection threshold.
Our preliminary asteroseismic analysis of J0840 yields a stellar mass $M=1.14\pm 0.01\,M_{\odot}$, hydrogen and helium envelope masses of $M_H = 5.8 \times 10^{-7}\,M_{\odot}$ and $M_{He}=4.5 \times 10^{-4}\,M_{\odot}$, and an expected core crystallized mass ratio of 50-70\%.
J1053, J1554, and J2038 have masses in the range $0.84-0.91 M_\odot$ and are expected to have a CO core; 
however, the core of J0840 could consist of highly crystallized CO or ONeMg given its high mass.
These newly discovered massive pulsators represent a significant increase in the number of known ZZ Ceti white dwarfs with mass
$M > 0.85\,M_\odot$, and detailed asteroseismic modeling of J0840 will allow for significant tests of crystallization
theory in CO and ONeMg core white dwarfs.
\end{abstract}

\begin{keywords}
white dwarfs, stellar pulsations
\end{keywords}

\begin{table*}
  \captionsetup{width=.75\textwidth}
  \caption{Observational properties of our WD sample.}
  \begin{tabular}{cccccccc}
   \hline
   SDSS & $g$   & S/N & $T_\textrm{\tiny{ef}f}$ & $\log{g}$ & Mass        & Period     & Amplitude\textsuperscript{1} \\
        & (mag) &     & (K)                     &           & (M$_\odot$) & (s)        &  (mma)    \\
   \hline
   J0116+3128 & $19.08$ & $19$ & $12\,210\pm370$ & $8.64\pm0.08$ & $1.01\pm0.05$ & - & $<7.4$ \vspace{5pt} \\ 
   J0446$-$0441 & $19.56$ & $17$ & $11\,830\pm380$ & $8.57\pm0.09$ & $0.97\pm0.06$ & - & $<21.6$ \vspace{5pt} \\  
   J0520+1710 & $19.15$ & $25$ & $12\,030\pm310$ & $8.78\pm0.07$ & $1.09\pm0.04$ & - & $<8.8$ \vspace{5pt} \\
   J0727+4036 & $18.10$ & $38$ & $12\,350\pm340$ & $9.01\pm0.07$ & $1.20\pm0.03$ & - & $<9.4$ \vspace{5pt} \\
   J0822+0824 & $18.12$ & $23$ & $11\,290\pm230$ & $8.47\pm0.07$ & $0.90\pm0.05$ & - & $<7.5$ \vspace{5pt} \\
   \textbf{J0840+5222} & $18.24$ & $36$ & $12\,160\pm320$ & $8.93\pm0.07$ & $1.16\pm0.03$ & $326.6\pm1.3$ & $7.1\pm1.0$\\
    & & & & & & $172.7\pm0.4$ & $6.2\pm1.0$\\
    & & & & & & $797.4\pm8.0$ & $6.3\pm1.0$ \vspace{5pt} \\
   J0904+3703 & $19.09$ & $18$ & $11\,800\pm320$ & $8.45\pm0.08$ & $0.89\pm0.05$ & - & $<6.7$ \vspace{5pt} \\
   J0942+1803 & $18.17$ & $25$ & $11\,380\pm210$ & $8.49\pm0.06$ & $0.91\pm0.04$ & - & $<4.5$ \vspace{5pt} \\
   \textbf{J1015+2340} & $18.67$ & $14$ & $11\,320\pm300$ & $8.44\pm0.10$ & $0.88\pm0.06$ & $498.5\pm4.9$ & $15.7\pm2.3$ \vspace{5pt} \\
   J1053+6347 & $18.65$ & $16$ & $12\,590\pm450$ & $8.64\pm0.09$ & $1.01\pm0.05$ & - & $<7.3$ \vspace{5pt} \\
   \textbf{J1554+2410} & $17.55$ & $27$ & $11\,470\pm230$ & $8.49\pm0.07$ & $0.91\pm0.04$ & $673.6\pm2.2$ & $17.9\pm1.1$ \vspace{5pt} \\
   J1655+2533 & $16.94$ & $34$ & $11\,060\pm170$ & $9.20\pm0.06$ & $1.27\pm0.02$ & - & $<2.5$ \vspace{5pt} \\
   \textbf{J2038+7710} & $19.05$ & $20$ & $11\,940\pm310$ & $8.38\pm0.08$ & $0.84\pm0.05$ & $203.7\pm0.1$ & $16.3\pm1.3$ \vspace{5pt} \\
   \hline
  \end{tabular}
  \label{tab:table1}
\end{table*}

\begin{table*}
  \captionsetup{width=.75\textwidth}
  \caption{Journal of observations for the thirteen ZZ Ceti candidates presented in this report. $\Delta{t}$ is the total integration time of the observations
           and $t_{\textrm{exp}}$ is the exposure time of each individual frame.}
  \begin{tabular}{ccccccc}
   \hline
   SDSS & Instrument (Telescope) & Filter & Date & $t_{\textrm{exp}}$ & $\Delta{t}$ & No. of points \\
        &                        &        &      & (s)                & (h) &                       \\
   \hline
   J011647.94+312845.7 & Agile (APO 3.5m) & BG40 & 2015 Oct 12 & 45, 60 & 1.63 & 117 \\
   (J0116+3128) & ProEM (McDonald 3.5m) & BG40 & 2014 Oct 30 & 5 & 3.81 & 2744 \\
    & & & & & \\
   J044628.66$-$044125.5 & ProEM (McDonald 2.1m) & BG40 & 2015 Feb 02 & 30 & 3.88 & 466 \\
   (J0446$-$0441) & Agile (APO 3.5m) & BG40 & 2015 Oct 12 & 45 & 1.59 & 127 \\
    & & & & & \\
   J052016.37+171003.0 & Agile (APO 3.5m) & BG40 & 2014 Jan 28 & 30 & 0.93 & 112 \\
   (J0520+1710) & ProEM (McDonald 2.1m) & BG40 & 2014 Oct 02 & 30 & 3.35 & 402 \\
                & ProEM (McDonald 2.1m) & BG40 & 2014 Oct 03 & 30 & 3.53 & 423 \\
                & ProEM (McDonald 2.1m) & BG40 & 2014 Oct 29 & 10 & 3.92 & 1411 \\
                & ProEM (McDonald 2.1m) & BG40 & 2014 Oct 30 & 5 & 2.70 & 1942 \\
                & GMOS-N (Gemini 8.1m) & \textit{g} & 2015 Feb 02 & 10 & 1.5 & 200 \\
    & & & & & \\
   J072724.66+403622.0 & Agile (APO 3.5m) & BG40 & 2016 Apr 05 & 40, 60 & 1.53 & 132 \\
   (J0727+4036) & & & & & \\
    & & & & & \\
   J082239.43+082436.7 & Agile (APO 3.5m) & BG40 & 2014 Jan 28 & 30 & 1.01 & 121 \\
   (J0822+0824) & & & & & \\
    & & & & & \\
   J084021.23+522217.4 & Agile (APO 3.5m) & BG40 & 2016 Jan 15 & 45 & 0.61 & 49 \\
   (J0840+5222) & Agile (APO 3.5m) & BG40 & 2016 Apr 04 & 40 & 1.79 & 161 \\
                & ProEM (McDonald 2.1m) & BG40 & 2016 May 04 & 10 & 3.10 & 1117 \\
                & ProEM (McDonald 2.1m) & BG40 & 2016 May 05 & 10 & 1.25 & 451 \\
    & & & & & \\
   J090459.26+370344.4 & ProEM (McDonald 2.1m) & BG40 & 2016 Jan 13 & 10 & 3.06 & 1102 \\
   (J0904+3703) & Agile (APO 3.5m) & BG40 & 2016 Jan 15 & 45 & 1.76 & 141 \\
    & & & & & \\
   J094255.02+180328.6 & GMOS-N (Gemini 8.1m) & \textit{g} & 2015 Apr 09 & 10 & 0.36 & 132 \\
   (J0942+1803) & GMOS-N (Gemini 8.1m) & \textit{g} & 2015 Apr 30 & 10 & 0.30 & 109 \\
                & GMOS-N (Gemini 8.1m) & \textit{g} & 2015 May 18 & 10 & 0.20 & 72 \\
    & & & & & \\
   J101540.14+234047.4 & GMOS-N (Gemini 8.1m) & \textit{g} & 2015 Mar 03 & 10 & 0.45 & 163 \\
   (J1015+2340) & & & & & \\
    & & & & & \\
   J105331.46+634720.9 & Agile (APO 3.5m) & BG40 & 2016 Jan 15 & 45 & 1.68 & 134 \\
   (J1053+6347) & & & & & \\
    & & & & & \\
   J155438.35+241032.6 & GMOS-N (Gemini 8.1m) & \textit{g} & 2015 Mar 16 & 10 & 0.08 & 27 \\
   (J1554+2410) & Agile (APO 3.5m) & BG40 & 2015 Apr 09 & 45 & 1.54 & 123 \\
                & GMOS-N (Gemini 8.1m) & \textit{g} & 2015 May 30 & 10 & 0.11 & 39 \\
    & & & & & \\
   J165538.93+253346.0 & GMOS-N (Gemini 8.1m) & \textit{g} & 2015 Apr 17 & 10 & 0.09 & 32 \\
   (J1655+2533) & ProEM (McDonald 2.1m) & BG40 & 2015 Aug 13 & 5 & 3.77 & 2711 \\
    & & & & & \\
   J203857.52+771054.6 & ProEM (McDonald 2.1m) & BG40 & 2014 Aug 04 & 25 & 3.95 & 569 \\
   (J2038+7710) & Agile (APO 3.5m) & BG40 & 2014 Aug 24 & 40 & 1.12 & 101 \\
                & ProEM (McDonald 2.1m) & BG40 & 2014 Sep 02 & 20 & 4.29 & 773 \\
    & & & & & \\
   \hline
  \end{tabular}
  \label{tab:table2}
\end{table*}

\section{Introduction}
White dwarfs (WDs) are the inert remnants of stars with a Zero Age Main Sequence (ZAMS) mass of less than $\approx8\,M_\odot$. 
With nuclear burning having ceased, WDs radiate away their energy and cool as a result.
As hydrogen atmosphere (DA) WDs age and cool, they evolve through the ZZ Ceti instability strip wherein they become pulsationally unstable.
The subsequent $g$-mode oscillations are excited by driving in the partial ionization zone of hydrogen in the atmosphere of the WD
\citep{fontaine2008,winget2008}.
A detailed pulsational analysis of these modes provides stringent constraints on the stellar mass and the thickness of the
surface hydrogen layer \citep{bischoff2014,giammichele2016}.

The extreme pressure and density present in cool WDs induces crystallization as thermal energy is lost \citep{kirshnitz1960,abriksov1960,salpeter1961} and 
this crystallization releases latent heat which significantly slows the WD cooling rate \citep{vHorn1968}. 
\citet{segretain1994} show that central crystallization in a WD releases enough energy to lengthen the cooling time by several Gyr.
Crystallization also affects the pulsations \citep{hansen1979}.
However, only high mass WDs have significantly crystallized cores while they are in the ZZ Ceti instability strip (Lamb \& Van Horn 1975).
Motivated by the discovery of pulsations in the massive white dwarf star BPM 37093 \citep{kanaan1992}, \citet{winget1997} show that the mean period spacing of radial overtones grows as the crystallized mass ratio increases.  
In the first applications of crystallization theory, \citet{montgomery1999} and \citet{metcalfe2004} obtain best-fit solutions to the pulsation spectrum of BPM 37093 
that indicate it is of mass $M \lesssim 1.1\,M_\odot$ with a crystallized mass ratio of $\approx90\%$.
In an independent analysis, \citet{brassard2005} conclude that the crystallized mass ratio lies between 32\% and 82\% given the unknown chemical composition of the core.

Further evidence for core crystallization in cool WDs comes from \citet{winget2009}, whose analysis of the luminosity function and color-magnitude diagram of the globular cluster NGC 6397 
provides strong evidence for a first-order phase transition and the release of latent heat, which are central aspects of crystallization theory
\citep{vHorn1968}.
However, crystallization theory has yet to be thoroughly tested largely due to the lack of a large sample of high mass ($M > 0.8\,M_\odot$), variable DA WDs (or DAVs).
The mass distribution of DA WDs peaks at $\approx 0.6\,M_\odot$, with a tail toward higher masses
\citep{liebert2005,kepler2007,tremblay2011,kleinman2013,kepler2015}.
Given the historically small number of known WDs, the number of massive DAVs has only recently begun to grow significantly. 
This has led to difficulties to identify WDs on the high mass end of the ZZ Ceti instability strip.
In fact, the ultramassive regime ($M \ge 1.1\,M_\odot$) for DAVs, until now, was populated solely by BPM 37093 and GD 518 \citep{hermes2013}.

\footnotetext[1]{1 mma = 0.1\% relative amplitude}

The Sloan Digital Sky Survey (SDSS) has increased the number of spectroscopically confirmed WDs to about 30,000
\citep{harris2003,kleinman2004,eisenstein2006,kleinman2013,kepler2015,kepler2016}.
Hence, the SDSS WD catalogs present an unprecedented opportunity to discover massive pulsating WDs
and to eventually carry out rigorous tests of crystallization theory, which served as an impetus for this work.
\citet{mukadam2004}, \citet{kepler2005}, \citet{mullally2005}, \citet{castanheira2006}, \citet{castanheira2009}, \citet{kepler2012}, and \citet{castanheira2013} have used the SDSS data
to search for DAVs, including massive ones. Currently, there are about 200 DAVs known.
\citet{castanheira2013} studied the ensemble properties of high mass DAVs 
and found evidence for a bimodal period distribution with no dominant pulsation
periods near $\approx$500 s, which may be evidence of a mode selection mechanism.
\citet{romero2013} analyzed the pulsation profiles of 42 high mass DAVs with $1.05\,M_\odot > M > 0.72\,M_\odot$ and found that a crystallized interior yields best-fitting solutions
for 15 stars. They also conclude that the mass of the hydrogen envelope in these stars ranges from $10^{-4}$ to $10^{-10}\,M_*$.
The recent discovery of the most massive ($M = 1.20 \pm 0.03\,M_\odot$) DAV, GD 518, by \citet{hermes2013} marks the beginning of the population of the extremely high mass end of the ZZ Ceti instability strip.
Such objects are likely the remnants of stars with a ZAMS mass of $\gtrapprox 7\,M_\odot$ and could contain ONe or ONeMg cores
as a result of carbon burning.
Further populating the high mass end of the ZZ Ceti instability strip will allow the ensemble characteristics of WDs in this regime to be analyzed for the first time.

In this paper, we present results from multiple observations conducted on DA WDs selected from the SDSS Data Release 7 \citep{kleinman2013} and Data Release 10
with the aim of discovering high mass ($M > 0.8\,M_\odot$) DAVs. We report the successful detection of pulsations in four of our targets, including the most massive DAV in it, J0840, which is the second most massive DAV discovered to-date.
In Section 2, we discuss the sample selection applied in this study.
In Sections 3 and 4, we discuss our observations and analysis.
In Section 5, we discuss the characteristics of our sample and conclude.

\begin{figure*}
 \centering
 \includegraphics[scale=0.75]{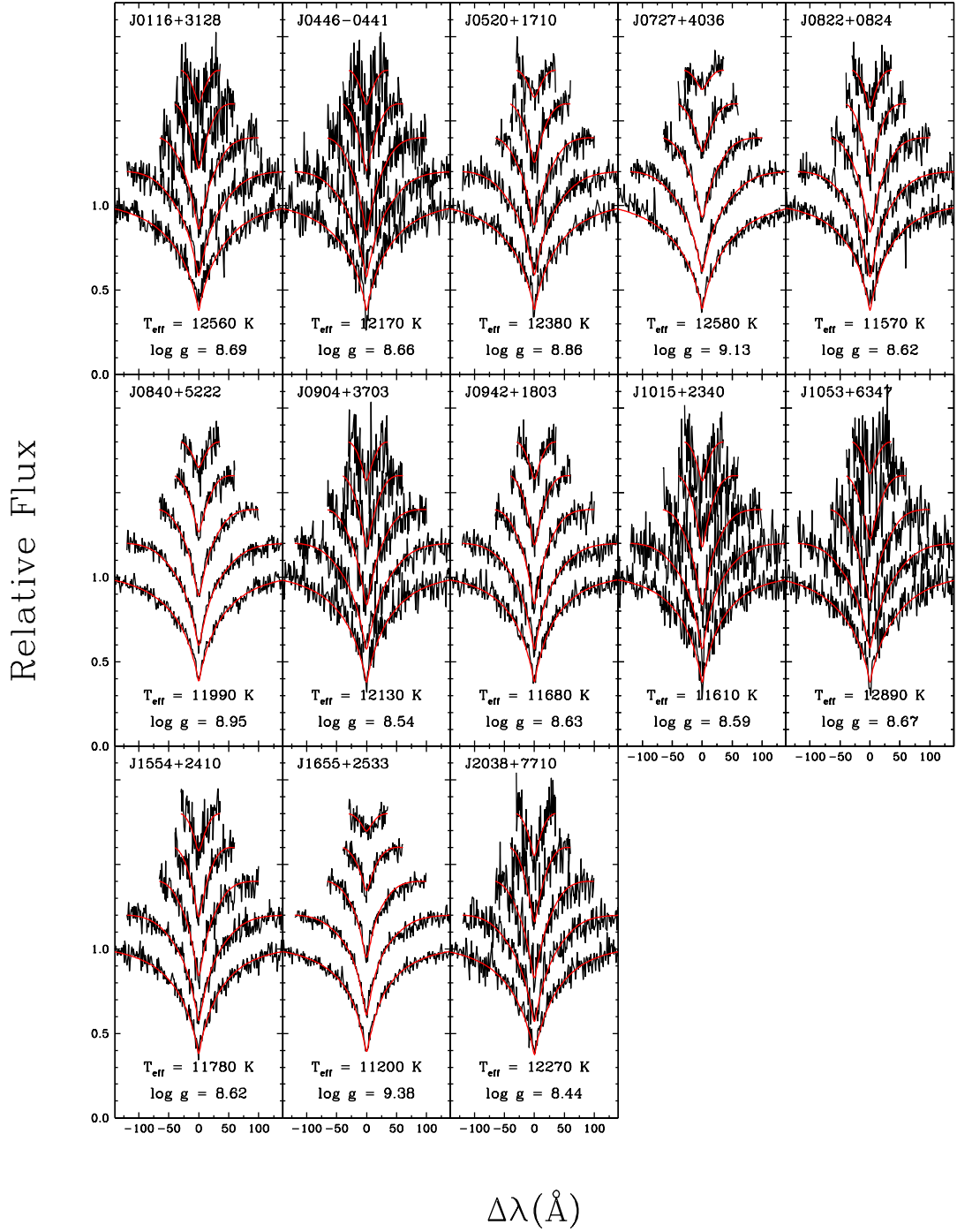}
 \caption{1D model astmosphere fits to the normalized Balmer line profiles of our targets.
          The best-fit parameters for each star are given in each panel.}
 \label{fig:p1}
\end{figure*}

\section{Sample Selection}

Our sample of targets consists of SDSS Data Release 7 (DR7) and Data Release 10 (DR10) WDs whose best-fitting atmospheric parameters place them within the empirical ZZ Ceti instability strip.

We select all targets with $T_\textrm{\tiny{ef}f} = 10,000 - 13,000$ K and $\log{g} > 8.5$ from the SDSS DR7 White Dwarf Catalog of
\citet{kleinman2013} as objects of interest.
For each of the 389 objects obtained in this first cut, we fit the normalized Balmer lines up to H$\epsilon$ of the SDSS spectra using the
procedures described in \citet{gianninas2011}.
We identify twelve targets within the DR7 sample whose $T_\textrm{\tiny{ef}f}$ and $\log{g}$ are within the empirical boundaries of the ZZ Ceti instability strip \citep{gianninas2011} given the estimated errors. 
\citet{kilic2015} photometrically identify one of these targets, J1529+2928, as a massive white dwarf with a dark spot. Hence, J1529+2928 is excluded from the following discussion.

We also obtained spectral fits to $\approx$ 6,000 DR10 WD spectra (identified by one of the authors, CAP) to search for additional targets.
Given the number of potential targets in the DR10 sample, we only select the brightest and most massive objects ($g < 18.5$ mag, $M > 1.05\,M_\odot$) for follow-up observations. 
We identify four targets matching these criterion and were able to observe two (J0727 and J0840).

Figure \ref{fig:p1} presents our best fits to the normalized Balmer line profiles using ML2/$\alpha = 0.8$ 1D model atmospheres for
our 13 massive DAV candidates. Table \ref{tab:table1} presents the best-fit parameters for these targets, including the 3D atmospheric
corrections from \citet{tremblay2013} and the average signal-to-noise ratio (S/N) of the SDSS spectrum. We computed the mass of each target using DA WD cooling models described in \citet{fontaine2001}. As discussed in \citet{gianninas2005}, since SDSS spectra are obtained over a set exposure time the S/N is significantly lower for fainter stars. For objects with a signal-to-noise ratio of $\sim 20$ (which is representative of our sample) the errors in effective temperature and surface gravity are as high as $400$ K and 0.1, respectively.  For such low S/N spectra this results in some ambiguity in selecting stars within the instability strip, especially near the edges of the strip. Nevertheless, we are confident that the solutions presented in Table \ref{tab:table1} are accurate, but not precise, resulting in uncertainties of up to 7\% in our mass estimates. We discuss this in more detail in Section 4.

\section{Observations}

We obtained follow-up time-series photometric data on the Gemini-North 8m telescope, the ARC 3.5m telescope at Apache Point Observatory (APO), and the Otto Struve 2.1m telescope at McDonald Observatory.

We aquired high speed photometry of ten objects using the ARC 3.5m telescope with the Agile frame transfer CCD with the BG40 filter.
Exposure times ranged from 30 to 60 s depending on the conditions and target brightness with uninterrupted integration times ranging from 0.6 to 1.8 hours.
We used the slow read-out setting and binned the CCD by 2 $\times$ 2, which resulted in a plate scale of 0.258 arcsec pixel$^{-1}$. 

We aquired high speed photometry of seven objects using the 2.1m Otto Struve telescope with the ProEM camera and the BG40 filter.
Exposure times ranged from 5 to 25 s depending on the conditions and target brightness with total integration times on the order of 3 to 4 hours.
We binned the CCD by 4 $\times$ 4, which resulted in a plate scale of 0.36 arcsec pixel$^{-1}$.

We observed five objects using the 8m Gemini-North telescope with
the Gemini Multi-Object Spectrograph (GMOS) as part of the queue program GN-2015A-Q-86.
We obtained time-series photometry for each of these WDs with 10 s exposures through an SDSS-$g$ filter.
We binned the CCD by 4 $\times$ 4, which yielded a read-out time and telescope overhead of $\approx$15 s and a plate scale of 0.29 arcsec pixel$^{-1}$.
Given the queue program, some of our targets were observed for less than 30 min, which is sufficient to confirm relatively high amplitude
pulsations, as in J1554. However, the total integration times were insufficient to detect lower amplitude (and possibly
longer period) pulsations in some of the other Gemini targets.
Table \ref{tab:table2} presents the journal of observations.

For each object, we obtain bias and flat field images and dark frames.
We reduced the GMOS data using the standard Gemini GMOS routines under the Image Reduction and Analysis Facility (IRAF).
We reduced the Agile and ProEM data using reduction routines in the IRAF \textit{imred} package.
We conduct aperture photometry on each object and nearby bright comparison stars in the images.
We use the IRAF \textit{digiphot} for aperture photometry on the GMOS and Agile data, and the external IRAF package \textit{ccd\_hsp}
\citep{kanaan2002} for aperture photometry on the ProEM data.
To correct for transparency variations, we divide the sky-subtracted light curves by the weighted sum of the light curves of the
nearest bright comparison stars in the field for each object.
We fit a third-order polynomial to each calibrated light curve to remove the low frequency signal ($P > 2000$ s) associated with a time varying transparency.
We note that our Fourier analysis only detects significant periods in the range of 150 s to 1000 s and thus the removal of such low frequency noise does not affect our conclusions.
We compute the discrete Fourier transform (DFT) of the calibrated, pre-whitened light curves up to the Nyquist frequency using the software package Period04 \citep{lenz2005}
and estimate the error associated with each period and amplitude using the Levenberg-Marquardt method as described in \citet{bevington1969}. 
We consider periods of amplitude greater than $4\langle{A}\rangle$ (as opposed to the less conservative $3\sigma$ threshold) to be a positive detection of pulsations, where $\langle{A}\rangle$ is the average amplitude of the DFT up to the Nyquist frequency.
\citet{breger1993} suggests that using a signal-to-noise amplitude ratio of $\approx{4}$ appropriately differentiates real and false detections in pulsation analysis. 

\begin{figure}
 \includegraphics[width=\columnwidth]{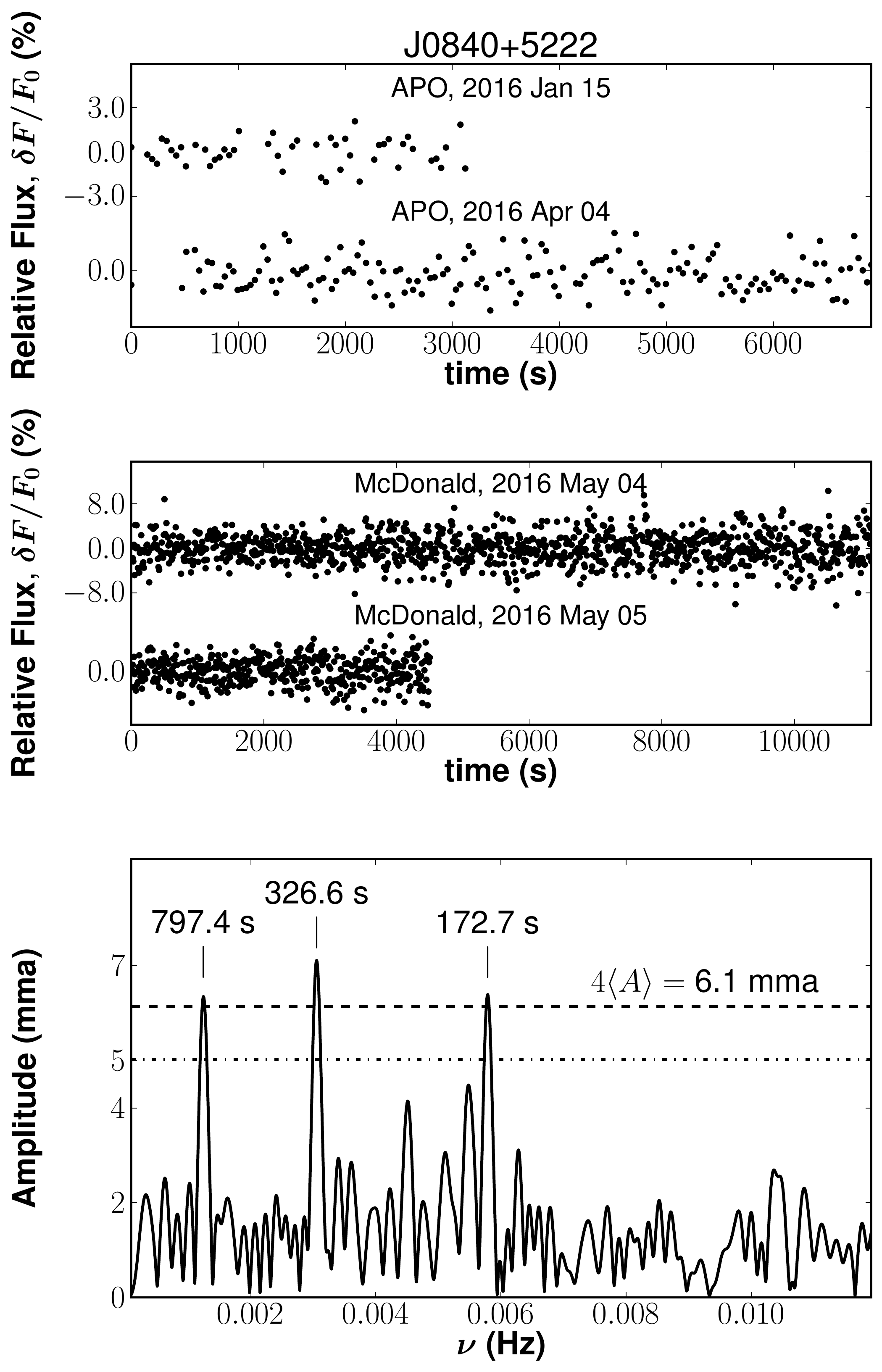}
 \caption{The light curves (top panels) and discrete Fourier transform (lower panel) from the longest integration time observations of SDSS J0840. In the lower panel, we indicate the $4\langle{A}\rangle$ (dashed line) and $3\sigma$ (dash-dotted line) detection limits.}
 \label{fig:p2}
\end{figure}

\begin{figure}
 \includegraphics[width=\columnwidth]{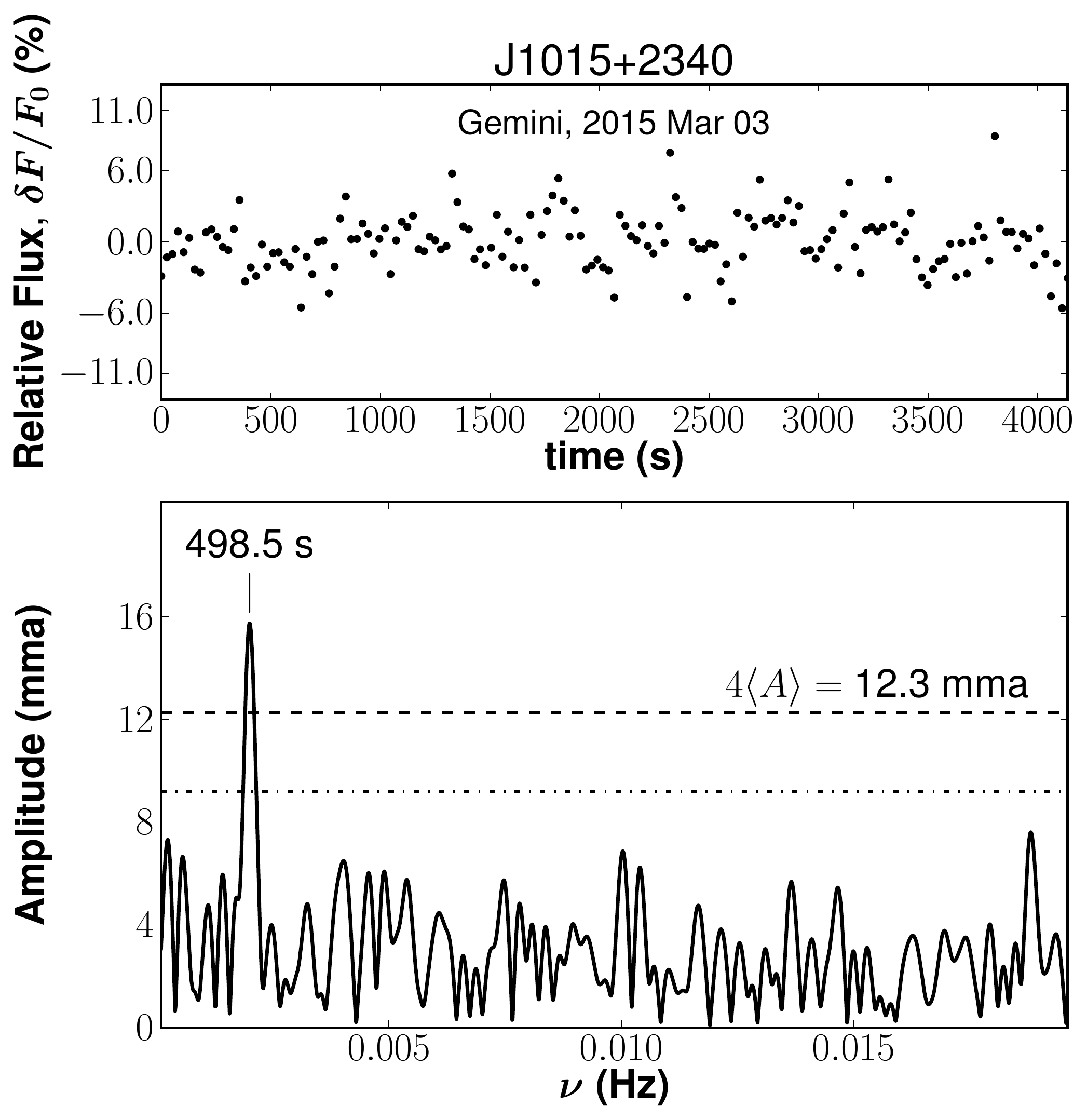}
 \caption{The light curve (top panel) and discrete Fourier transform (lower panel) of SDSS J1015. In the lower panel, we indicate the $4\langle{A}\rangle$ (dashed line) and $3\sigma$ (dash-dotted line) detection limits.}
 \label{fig:p3}
\end{figure}

\begin{figure}
 \includegraphics[width=\columnwidth]{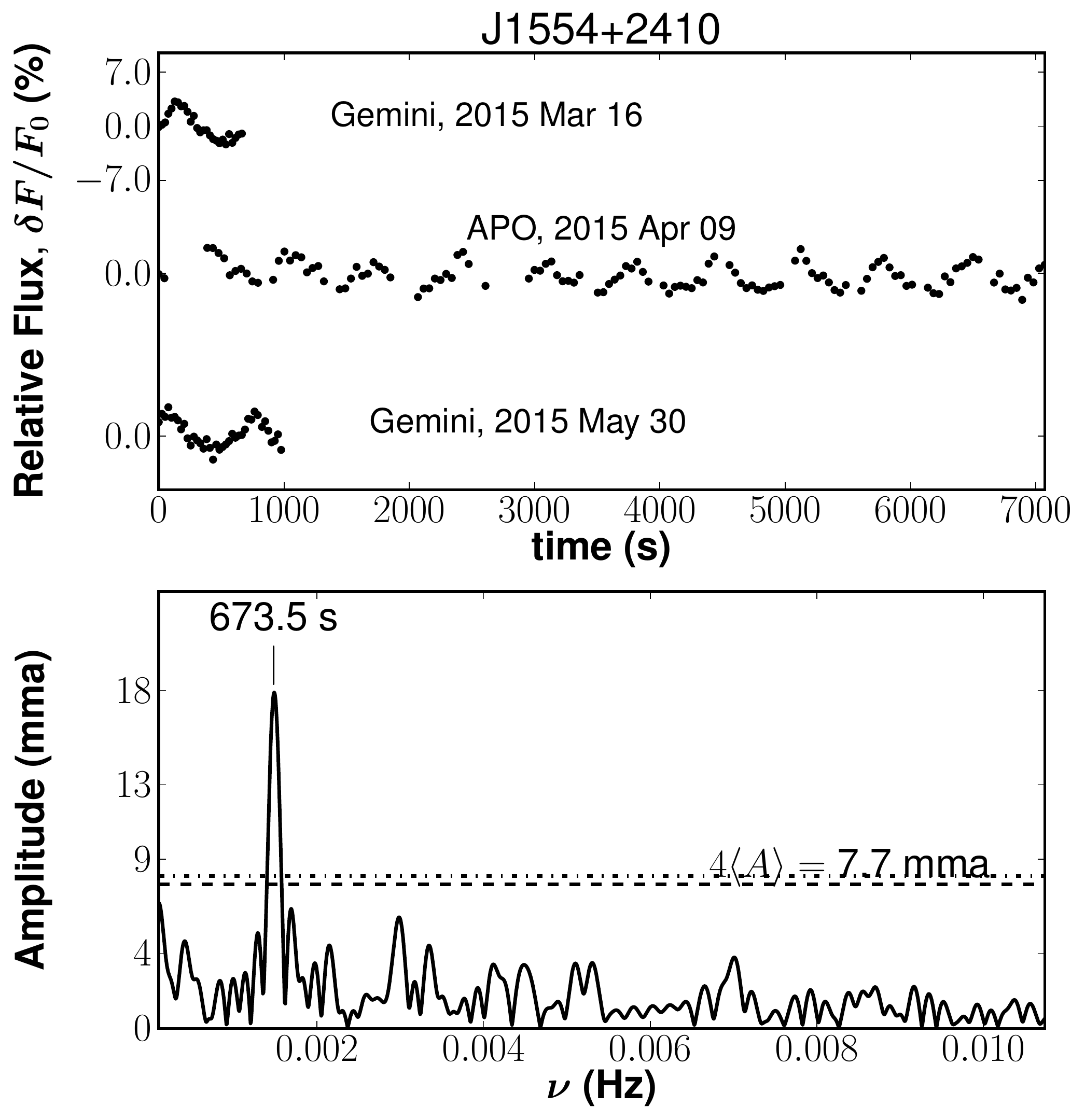}
 \caption{The light curves (top panel) and discrete Fourier transform (lower panel) from the longest integration time observations of SDSS J1554. In the lower panel, we indicate the $4\langle{A}\rangle$ (dashed line) and $3\sigma$ (dash-dotted line) detection limits.}
 \label{fig:p4}
\end{figure}

\begin{figure}
 \includegraphics[width=\columnwidth]{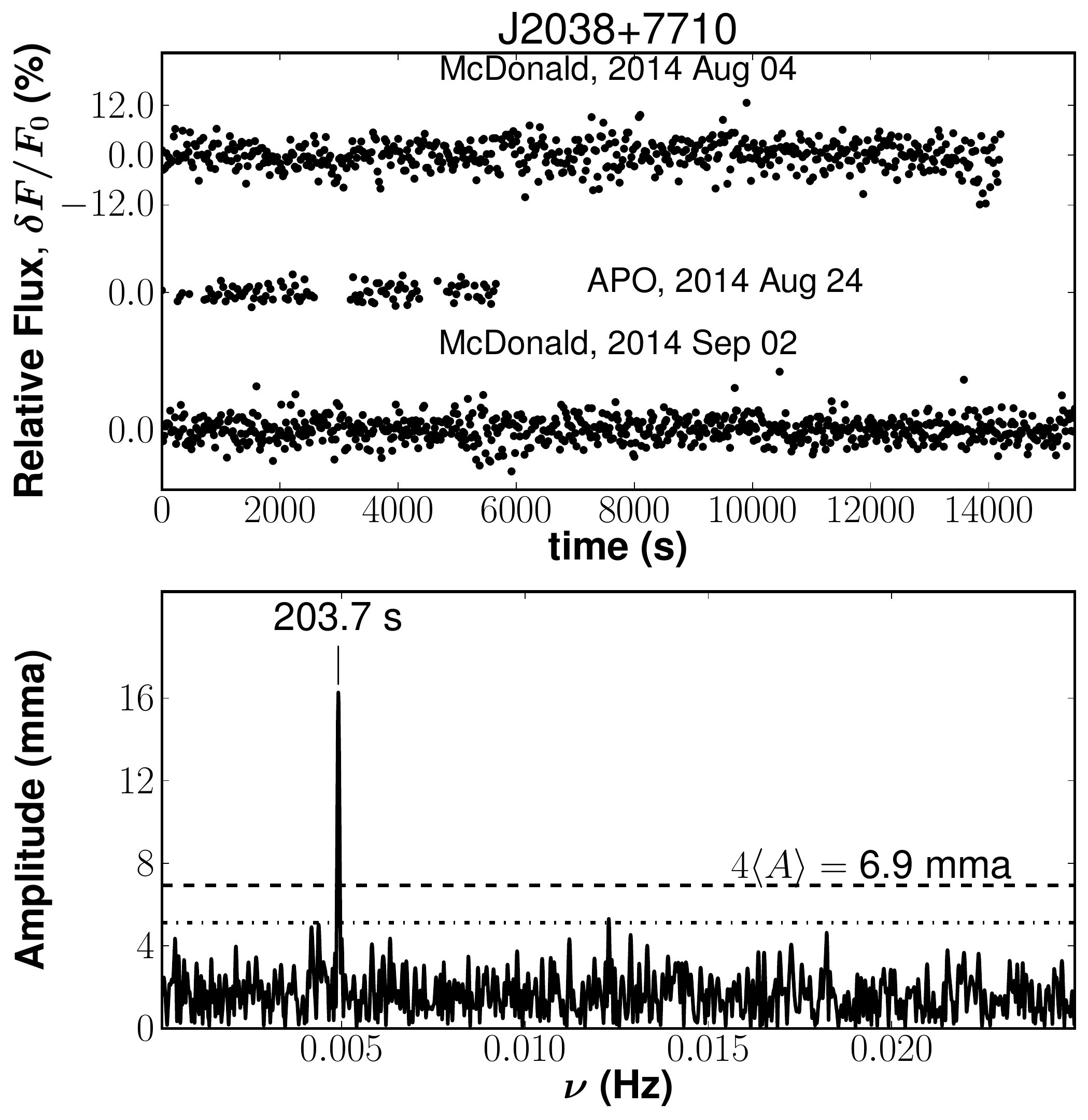}
 \caption{The light curves (top panels) and discrete Fourier transform (lower panel) from the longest integration time observations of SDSS J2038. In the lower panel, we indicate the $4\langle{A}\rangle$ (dashed line) and $3\sigma$ (dash-dotted line) detection limits.}
 \label{fig:p5}
\end{figure}

\section{Analysis}
Here we describe the properties of each discovered DAV including the period(s) and amplitude(s) detected in our Fourier analysis.
We then characterize the targets in our sample for which we did not detect pulsations and discuss possible reasons for the large number of non-DAVs in our sample.
Lastly, we describe the results from our asteroseismic analysis of J0840.

\subsection{Newly Discovered DAVs}

\subsubsection{J0840}

J0840 is the most massive DAV in our sample with $M = 1.16\pm0.03\,M_\odot$, $T_{\textrm{ef{f}}} = 12\,160\pm320$ K, and $\log{g} = 8.93\pm0.07$.
Figure \ref{fig:p2} presents the APO and McDonald high speed photometry observations of J0840, along with the Discrete Fourier transform (DFT)
of the APO data from UT 2016 Apr 4. There are three significant frequencies, with the dominant period at $P = 326.6\pm1.3$ s with
$A = 7.1\pm1.0$ mma amplitude, and two other frequencies at $P = 797.4\pm8.0$ s and $P = 172.7\pm0.4$ s with amplitudes 
$A = 6.3\pm1.0$ mma and $A = 6.2\pm1.0$ mma respectively. Table \ref{tab:table3} presents the
periods and amplitudes of pulsation and the detection limits for each night of observations.

We confirm all three periods with the McDonald 2.1m telescope data from UT 2016 May 4 and 5, and also confirm the two shorter periods
($174.6\pm1.1$ s and $340.4\pm4.3$ s) with data from UT 2016 Jan 28. Given the different signal-to-noise ratios of the light curves
from each night, some of these modes fall below the $4\langle{A}\rangle$ limit, but they are persistent at both the APO
and McDonald data, and therefore must be real.
The computed amplitudes for the respective observations are consistent within the errors and thus these results do not indicate a 
modulation in the amplitude of these modes of pulsation.
Our best data (UT 2016 Apr 4 and May 4) suggest that the two shorter periods ($P_1 \approx 330$ s and $P_2 \approx 170$ s) may be overtones
as the frequencies are integer multiples of one another within the estimated errors.

\subsubsection{J1015}

J1015 is a DAV of mass $M = 0.88\pm0.06\,M_\odot$ with $T_{\textrm{ef{f}}} = 11\,320\pm300$ K and $\log{g} = 8.44\pm0.10$.
Figure \ref{fig:p3} shows the Gemini light curve of J1015 along with its DFT.
J1015 displays significant pulsations with period $P = 498.5\pm4.9$ s and amplitude $A = 15.7\pm2.3$ mma.
We note that the successful detection of a period of $P = 498.5\pm4.9$ s is contradictory to the suggestion made in \citet{castanheira2013} that the period distribution is bimodal and bereft of periods near $\approx 500$ s. Follow-up observations to verify the dominant pulsation period of J1015 are needed.

\subsubsection{J1554}

J1554 is a DAV of mass $M = 0.91\pm0.04\,M_\odot$ with $T_{\textrm{ef{f}}} = 11\,470\pm230$ K and $\log{g} = 8.49\pm0.07$.
Figure \ref{fig:p4} shows the Gemini and APO data on J1554 along with the DFT of the longest light curve from 
APO observations on UT 2015 Apr 9.
These data reveal a dominant pulsation mode at $P = 673.6\pm2.3$ s and amplitude $A = 17.9\pm1.1$ mma. On the other hand, the Gemini data show a significant peak at $P = 710.8\pm0.002$ s with an amplitude $A = 21.8\pm1.2$ mma.
However, given the brevity of the combined Gemini-North observations (the total integration time is less than $2000$ s),
the change in amplitude is likely not real.

To explore the effect of light curve gaps and noise on the resulting frequency power spectrum, we created synthetic light curves that emulate the 
observations of J1554 (Table \ref{tab:table2}, Figure \ref{fig:p4}) using a function of the form:
\begin{equation}
  A(t) = A_0\sin(2\pi{f_0}t) + N(\mu,\sigma),
\end{equation}
where $A_0$ is the amplitude of the pulsation in mma, $f_0$ is the frequency of the pulsation in Hz, $N(\mu,\sigma)$ (or the ``noise function") 
is a random number generator that samples a normal distibution with mean $\mu$ and standard deviation $\sigma$.
Note that we assume the presence of only one pulsation period (based on the APO observations, $A_0 = 17.9$ mma, $f_0 = 1\,487.8 \,\mu$Hz)
and that the noise is Gaussian.
We derive $\sigma$ from each respective pre-whitened light curve with the contribution from the pulsations subtracted.
Given the small number of data points in the Gemini observations, we run six initializations for values of 
$\sigma = 0.5\sigma_1, \sigma_2, \sigma_1, \textrm{ and } 2\sigma_2$ respectively (where $\sigma_1$ is the standard deviation from UT 2015 Apr 09 and $\sigma_2$ is the standard deviation from UT 2015 Mar 16 \& May 30).

\begin{table}
  \captionsetup{width=\columnwidth}
  \caption{Periods, amplitudes, and detection limits as determined for each night of observations for J0840.}
  \begin{tabular}{ccccc}
   \hline
   Date & Period & Amplitude   & $3\sigma$ & $4\langle{A}\rangle$ \\
        & (s) & (mma) &  (mma) & (mma) \\
   \hline
   2016 Jan 28 & $174.6\pm1.1$ & $6.7\pm1.4$ & $6.4$ & $8.9$ \\
               & $340.4\pm4.4$ & $5.5\pm1.4$ &  & \\
    & & & & \\
   2016 Apr 04 & $172.7\pm0.4$ & $6.2\pm1.0$ & $5.0$ & $6.1$ \\
               & $326.6\pm1.3$ & $7.1\pm1.0$ &  & \\  
               & $797.4\pm8.0$ & $6.3\pm1.0$ &  & \\
    & & & & \\
   2016 May 04 & $172.9\pm1.0$ & $8.1\pm1.1$ & $3.7$ & $5.5$ \\
               & $328.8\pm1.0$ & $5.8\pm1.1$ &  & \\  
               & $817.6\pm7.8$ & $4.5\pm1.1$ &  & \\
    & & & & \\
   2016 May 05 & $172.6\pm0.7$ & $7.7\pm1.5$ & $5.3$ & $7.8$ \\
               & $332.3\pm3.4$ & $6.4\pm1.5$ &  & \\  
               & $784.8\pm17.7$ & $6.0\pm1.5$ &  & \\
   \hline
 \end{tabular}
  \label{tab:table3}
\end{table}

\begin{table}
  \captionsetup{width=.75\textwidth}
  \caption{Ensemble characteristics of synthetic light curves of J1554. Note that $\sigma$ is the standard deviation of the ``noise function",
           $A_\textrm{min}$ is the smallest computed amplitude, $A_\textrm{max}$ is the largest computed amplitude, and $dA$ is the range of the set of solutions.
           Input parameters are $A_0 = 17.9$ mma, $f_0 = 1\,487.8 \,\mu$Hz, $\sigma_1=8.8\textrm{ mma }$, and $\sigma_2=7.0\textrm{ mma }$.}
  \begin{tabular}{ccccccc}
   \hline
   Data Set & $\sigma$ & $A_\textrm{min}$ & $A_\textrm{max}$ & $dA$ \\
            &          & (mma)            & (mma)            & (mma) \\
   \hline
   APO & $0.5\sigma_1$ & $17.5\pm0.6$ & $18.3\pm0.5$ & $0.8$ \\
       & $\sigma_2$ & $16.7\pm0.9$ & $19.6\pm1.0$ & $2.8$ \\  
       & $\sigma_1$ & $17.1\pm1.0$ & $20.2\pm1.1$ & $3.1$ \\ 
       & $2\sigma_2$ & $15.9\pm1.7$ & $21.5\pm1.8$ & $5.6$ \\
    & & & & \\
   Gemini & $0.5\sigma_1$ & $16.7\pm0.7$ & $18.9\pm0.8$ & $2.2$ \\
          & $\sigma_2$ & $16.5\pm1.1$ & $19.3\pm1.2$ & $2.8$ \\ 
          & $\sigma_1$ & $15.5\pm1.5$ & $20.0\pm1.5$ & $4.5$ \\
          & $2\sigma_2$ & $16.2\pm2.3$ & $25.0\pm2.6$ & $8.9$ \\
   \hline
 \end{tabular}
  \label{tab:table4}
\end{table}

We present the ensemble characteristics of our synthetic light curves for six initializations with $\sigma$ held constant in Table \ref{tab:table4}.
This analysis demonstrates that random noise and gaps results in a range of amplitude solutions of up to $8.9$ mma in the most extreme case of $\sigma=2\sigma_2$.
Even if we consider a modest noise contribution of $\sigma=\sigma_2$, the smallest amplitude computed from the APO data ($A_\textrm{min} = 16.7\pm0.9$ mma) and the largest 
amplitude computed from the Gemini data ($A_\textrm{max} = 19.3\pm1.2$ mma) are not consistent given the estimated errors.
We conclude that gaps and random noise, especially for small data sets, can result in amplitude solutions for two respective data sets that are inconsistent within the estimated errors 
despite the input amplitude remaining constant and thus we do not consider our observations of J1554 to be indicative of amplitude modulations.
There is no evidence of amplitude modulations in our other DAVs with multiple nights of data.
\citet{montgomeryANDodonoghue1999} show that nonlinear least squares errors can significantly underestimate the true error and should be treated as a lower limit, which further suggests that the inconsistency of 
amplitude and period solutions for J1554 is not strong evidence of modulations.

\begin{figure*}
 \includegraphics[width=\textwidth]{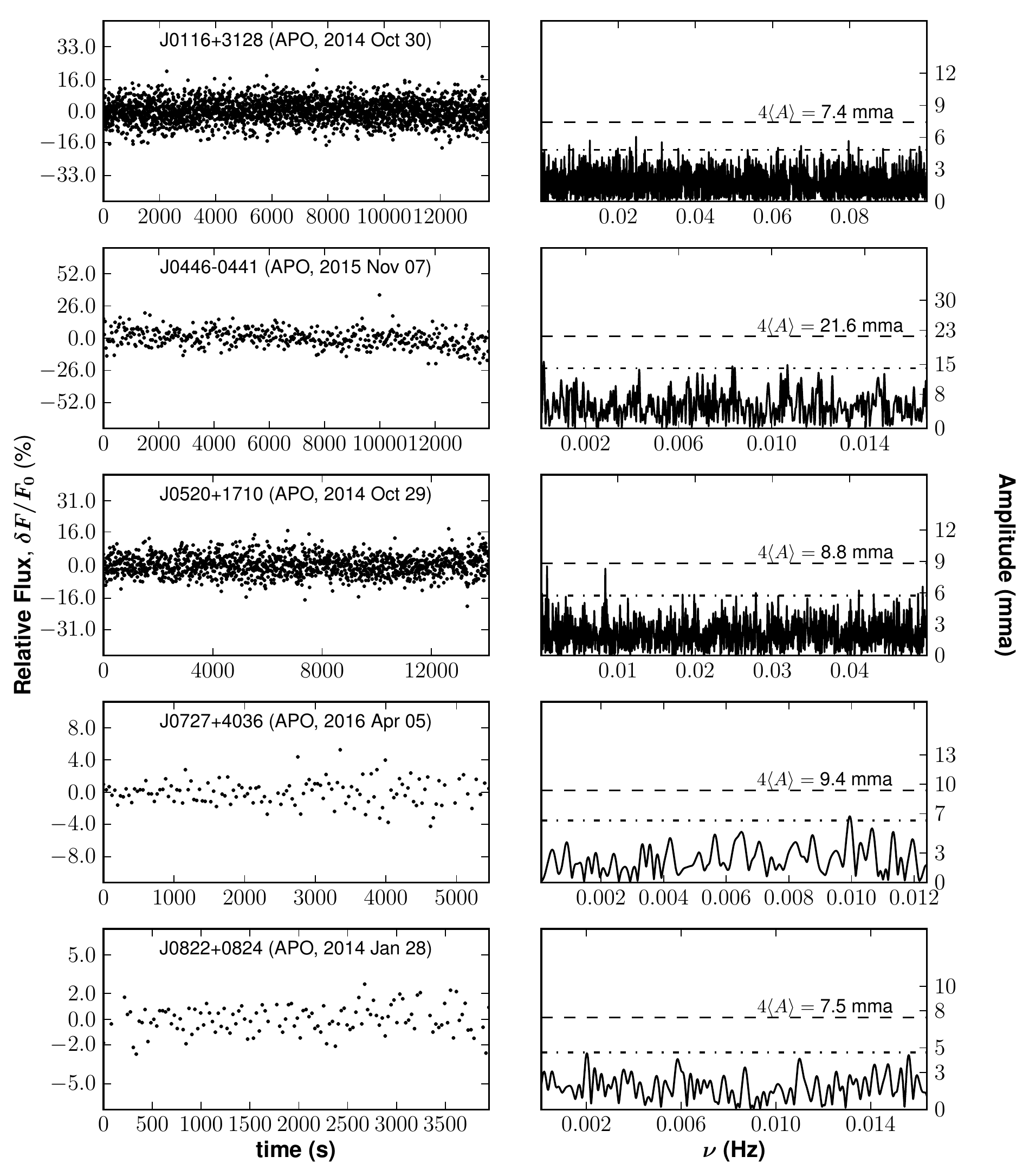}
 \caption{The light curve (left) and discrete Fourier transform (right) for the longest integration time observation for each respective WD in which we do not detect significant frequencies in the Fourier power spectrum. In the panels on the right, we indicate the $4\langle{A}\rangle$ (dashed line) and $3\sigma$ (dash-dotted line) detection limits.}
 \label{fig:p6}
\end{figure*}
\begin{figure*}
 \includegraphics[width=\textwidth]{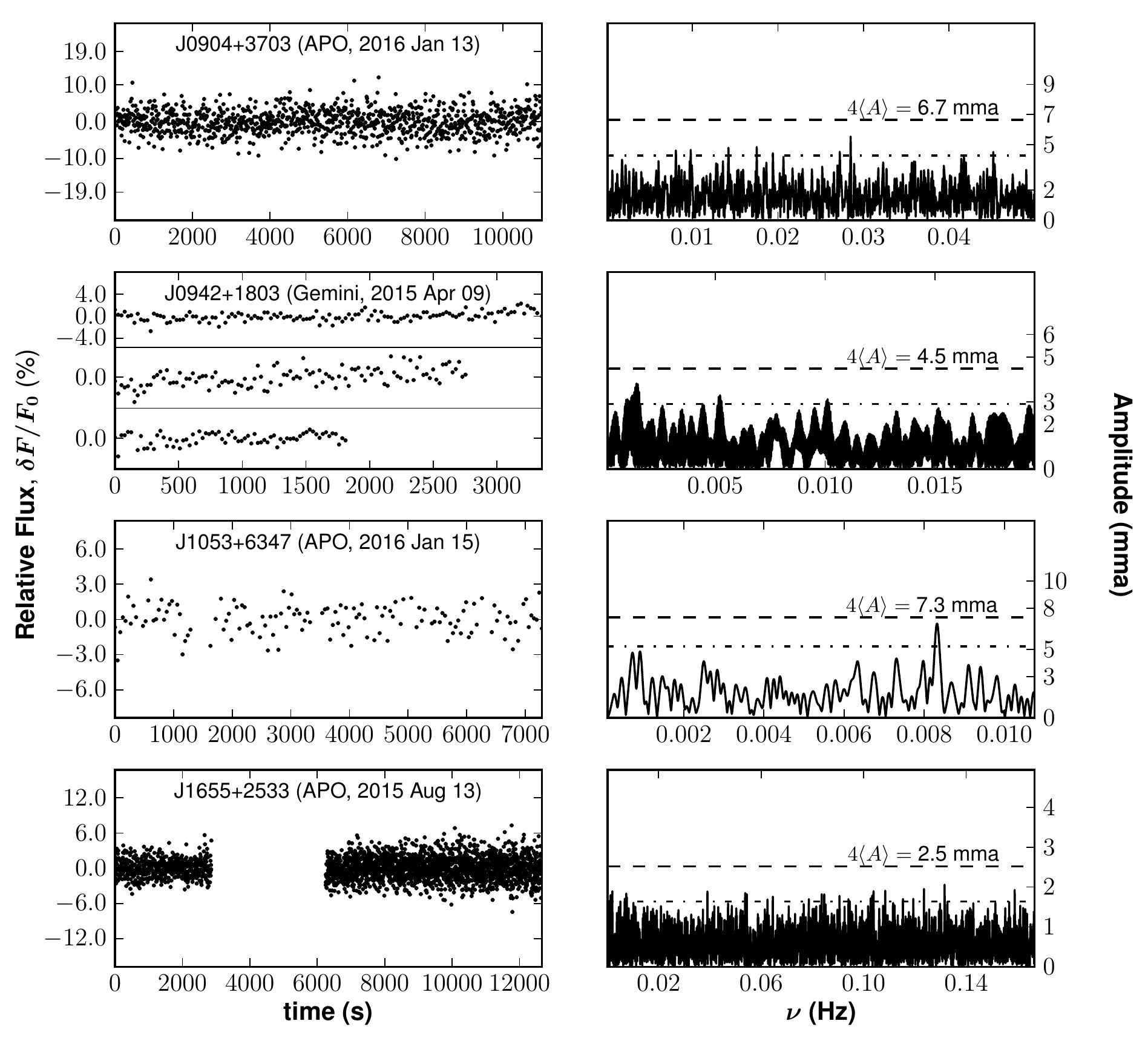}
 \caption{The light curve (left) and discrete Fourier transform (right) for the longest integration time observation for each respective WD in which we do not detect significant frequencies in the Fourier power spectrum. In the panels on the right, we indicate the $4\langle{A}\rangle$ (dashed line) and $3\sigma$ (dash-dotted line) detection limits}
 \label{fig:p7}
\end{figure*}

\begin{figure*}
 \centering
 \includegraphics[scale=0.55,angle=270]{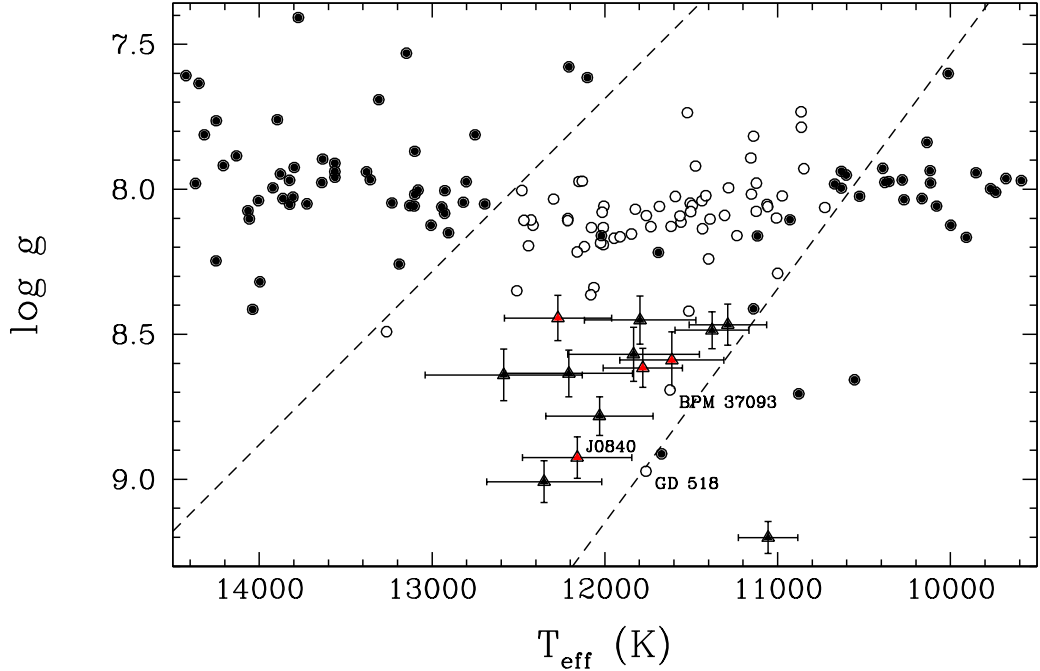}
 \caption{The 3D corrected atmospheric parameters of the newly discovered ZZ Ceti WDs (red triangles) and non-DAVs (black triangles) along
          with those of known ZZ Ceti WDs (open circles) and non-variable WDs (filled circles). The atmospheric parameters of our newly discovered DAVs are consistent with
          the empirical bounds of the ZZ Ceti strip (dashed lines) \citep{gianninas2011}. Previously known DAVs/non-DAVs include BPM 37093 \citep{kanaan1992} and GD 518 \citep{hermes2013} as well as DA WDs 
          from \citet{gianninas2011} and \citet{green2015}. }
 \label{fig:p8}
\end{figure*}

\begin{figure}
 \includegraphics[width=\columnwidth]{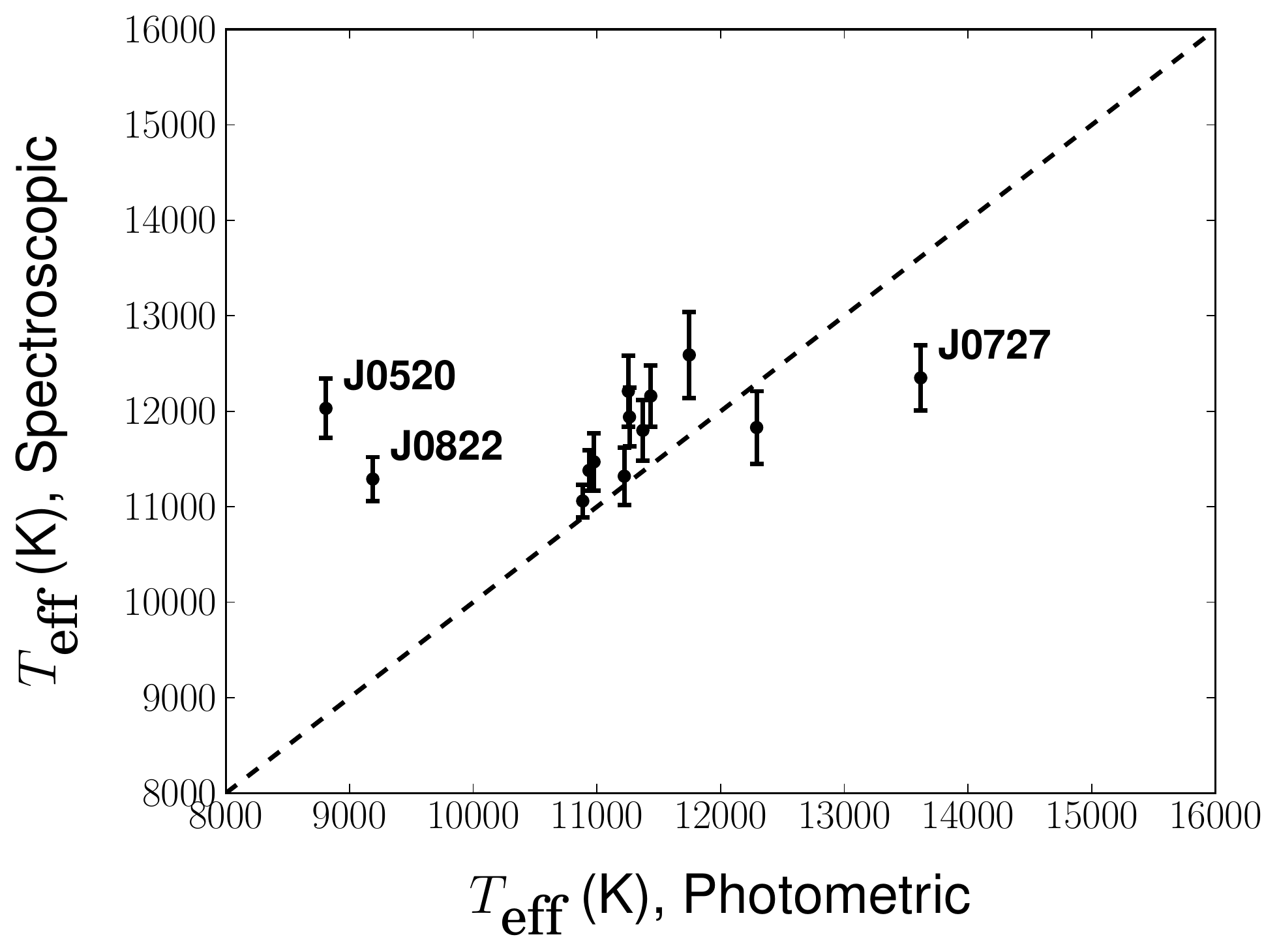}
 \caption{Comparison of the spectroscopic and photometric solutions for the effective temperature of all thirteen targets in our sample.
          The diagonal dashed line shows one-to-one correspondence.}
 \label{fig:p9}
\end{figure}

\subsubsection{J2038}

J2038 is a DAV of mass $M = 0.84\pm0.05\,M_\odot$ with $T_{\textrm{ef{f}}} = 11\,940\pm310$ K and $\log{g} = 8.38\pm0.08$.
Figure \ref{fig:p5} shows the McDonald and APO light curves of J2038 along with the DFT of longest light curve from McDonald observations
on UT 2014 Sep 2. J2038 displays significant pulsations with period $P = 203.7\pm0.1$ s and amplitude $A = 16.3\pm1.3$ mma.
We also compute the DFT using data from McDonald, 2014 Sep 02 (APO, 2014 Aug 24) and detect a period of $P = 203.8\pm0.2$ ($203.4\pm0.2$) s and amplitude of $A = 17.7\pm1.8$ ($18.5\pm1.1$) mma.
The computed amplitudes are consistent within the errors and thus these results do not indicate a variation in the amplitude of the dominant period of pulsation.

\subsection{Non-DAVs}

Figures \ref{fig:p6} and \ref{fig:p7} display light curves and DFTs of the longest integration time observations of the nine non-DAVs in
our sample.
Our $4\langle{A}\rangle$ detection limits for these objects ranges from 2.5 mma for J1655 to 21.6 mma for J0446.
We note that all but one of the objects (J1655) for which pulsations were not detected have spectroscopically determined atmospheric parameters that indicate they should undergo pulsations (Figure \ref{fig:p8}).
We computed the photometric temperature from the SDSS photometry following the procedure outlined by \citet{bergeron1997} and assuming $\log{g}=8.5$ for all thirteen targets.
We find good agreement between the photometric and spectroscopic solutions for all objects but J0520, J0727, and J0822 (see Figure \ref{fig:p9}). 
The photometric solutions for J0520 and J0822, respectively, are significantly cooler than the spectroscopic models and would place them outside of the ZZ Ceti instability strip.
However, the photometric temperature solution for J0727 (which is roughly $1\,000$ K hotter) is still within the bounds of the ZZ Ceti instability strip given its high surface gravity.

The rest of the non-variable objects are close to the red edge of the instability strip. Given the relatively low signal-to-noise ratio
of the SDSS spectra, the best-fitting temperature solutions for these stars are within 1-2$\sigma$ of the empirical red-edge, hence we
suspect that some of these stars might have already evolved outside of the instability strip. Another alternative is that
our $4\langle{A}\rangle$ detection limits are not stringent enough to detect low amplitude pulsations in these stars.
For example, we detect a single low amplitude period below the $4\langle{A}\rangle$ threshold but above the 3$\sigma$ level
for both J0727 and J1053. J0727 shows a dominant period at $P = 100.54\pm0.97$ s and amplitude $A = 6.66\pm1.76$ mma, whereas
J1053 shows a dominant period at $P = 120.21\pm0.25$ s and amplitude $A = 6.79\pm1.37$ mma. 
These periods lie within the observed range of periods for DAVs, which is roughly 100-2000 s.
J0520 and J0904 also showed a significant period above the $3\sigma$ threshold on one night (2014 Oct 29 and 2016 Jan 15, respectively); however, 
no subsequent detections were made in any of the other available light curves.
A better sampling rate and longer integration time would significantly improve the signal-to-noise in the DFT and may reveal these periods to be indicative of real pulsations.

Given the relatively short timespan of our observations of some of the non-DAVs and our detection limits, it is possible that several of these WDs may exhibit lower amplitude pulsations that would escape detection in our light curves.
For example, several DAVs presented in \citet{castanheira2013} and \citet{romero2013} have dominant periods of pulsation with amplitudes lower 
than 5 mma. In addition, J0840, a newly discovered DAV in our sample, is another example of a ZZ Ceti WD with two low amplitude pulsations
with $A < 7$ mma which is below the computed detection threshold for both J0116 and J0446.

\subsection{Preliminary Asteroseismic Analysis of J0840}
We used the three detected pulsation periods from Apr 04 (see Table \ref{tab:table3}) to perform a preliminary asteorseismological analysis. The DA WD models used in this work were generated using LPCODE evolutionary code (see \citet{renedo2010} for details). We employed WD model sequences with stellar mass between 1.024 $M_{\odot}$ and 1.15 $M_{\odot}$ and carbon-oxygen core. The sequences with stellar mass larger than 1.08 $M_{\odot}$ were obtained by artificially scaling the stellar mass from the $\sim 1M_{\odot}$ sequence at high effective temperatures, and the remaining ones are those presented in \citet{romero2013}. 

Residual nuclear burning was not considered for the massive sequences, since they have relatively thin hydrogen envelopes. Crystallization processes and the additional energy sources are included following the \citet{horowitz2010} phase diagram. Note that the effective temperature where crystallization begins increases with stellar mass \citep{romero2013}. 
For instance, crystallization starts at $14\,500$ K for $1.05\,M_{\odot}$ and at $17\,250$ K for $1.14\,M_{\odot}$.
Finally, we computed non-radial $g$-mode pulsations using the adiabatic version of the LP-PUL pulsation code described in \citet{corsico2006} (see \citet{romero2012,romero2013} for details), we consider $\ell=1,2$ modes.

Using the mass-radius relations from \citet{romero2012} and \citet{althaus2005} for the O/Ne core WDs, we determine a stellar mass of $1.13 \pm 0.05 M_{\odot}$. Note that the limiting mass for C/O core WDs is uncertain and it depends on the metallicity of the progenitor star. According to \citet{doherty2015}, for solar metallicity, the limiting mass is between $1.075 M_{\odot}$ and $ 1.158 M_{\odot}$.

To find the best fit model we minimize the quality function $S$:
\begin{equation}
S =\sqrt{ \sum^{N}_{i=1} \frac{[\Pi_k^{th} -\Pi_i^{obs}]^2 \times w_i }{\sum^N_{i=1}wi}}
\end{equation}

\noindent where $N$ is the number of observed modes and $w_i$ are the amplitudes.
First we consider only monopole modes. We obtained a solution characterized by a stellar mass $M=1.14 M_{\odot}$, $M_H = 5.837 \times 10^{-7} M_{\odot}$ and $M_{He}=4.455 \times 10^{-4} M_{\odot}$, $T_{eff} = 11\, 850$ K, with theoretical periods 171.862 s (l=1, k=3), 320.013 s (l=1, k=8) and 798.773 s (l=1, k= 23) and a value of $S=4.06$ s. Note that this model shows a lower effective temperature than the spectroscopic value, but is still in agreement considering the uncertainties. We note from our fits, that the dominant mode in determining the stellar mass is the mode $\sim 172$ s, which is also the more stable mode throughout the different nights. In this fit the mode 797.4 s is very well fitted but is also the mode with the larger uncertainty. 
Next, we include in our fit $\ell=2$ modes, and in addition we consider the uncertainties in the periods. We obtain a representative model characterized by the same stellar mass and hydrogen mass as our previous fit, but with $T_{\rm eff}=12\, 200$ K and theoretical periods 170.557 s (l=1, k=3), 326.562 s (l=2, k=15) and 804.645 s (l=2, k= 40).

It is important to note that with this set of periods, considering the uncertainties, the seismological effective temperature can vary from $11\, 850$ to $12\, 350$ K. On the other hand the stellar mass is more constrained to $M_* = 1.14 \pm 0.01 M_{\odot}$, in well agreement with the spectroscopic value. Within this effective temperature range, it is expected that 50-70 \% of the core mass to be crystallized. In this scenario the pulsation modes propogate in a small region of the star, located mainly in the envelope of the star.   

\section{Discussion and Conclusions}

We have confirmed pulsations in four DA WDs with $M > 0.84\,M_\odot$ including a $1.16\pm0.03\,M_\odot$ WD (J0840).
For the remaining nine targets in our sample we provide upper limits on their variability.

Our preliminary asteroseismic analysis of J0840 yields $M=1.14\pm 0.01\,M_{\odot}$,
$M_H = 5.8 \times 10^{-7}\,M_{\odot}$, $M_{He}=4.5 \times 10^{-4}\,M_{\odot}$, and an expected core crystallized mass ratio of 50-70\%.
We detected a period of $P = 498.5\pm4.9$ s in J1015. This result is contradictory to the suggestion made in \citet{castanheira2013} that the DAV period distribution is bimodal and bereft of periods near $\approx 500$ s. As only one night of data was available for J1015, follow-up observations are needed to verify this result.
The addition of these high mass DAVs nearly doubles the number of known ZZ Ceti WDs with $\log{g} > 8.5$ (Figure \ref{fig:p8}) which is a significant population in terms of probing stellar evolution and exploring crystallization theory.

At $M = 1.16\pm0.03\,M_\odot$, J0840 is of similar mass to the previously discovered ultramassive DAVs, BPM 37093 ($M \approx 1.10\,M_\odot$) and GD 518 ($M = 1.20\pm0.03\,M_\odot$).
We detect periods in J0840 ranging from roughly 180 to 800 s meanwhile pulsational periods of BPM 37093 lie in a narrow range of about 510 to 660 s \citep{kanaan2005} and those of GD 518 range from 425 to 595 s \citep{hermes2013}.
Our observed period range is consistent with the calculated period range for a $1.1\,M_\odot$ CO-core WD with $T_\textrm{ef{f}}=12,200$ K presented in Figure 9 of \citet{montgomery1999}, 
which shows the periods of $l$=2 modes for crystallized mass ratios ranging from 0 to 0.99. 

Previous studies of BPM 37093, the most extensively observed high mass DAV, have encountered difficulty with mode identification since its modes are both low
amplitude and undergo amplitude modulation \citep{kanaan2005}.
Furthermore, amplitudes reported for GD 518 range from roughly 1 to 4 mma \citep{hermes2013}. 
The dominant periods of pulsation for J0840 remained consistent within the errors between the four nights of available data over a period
of about 3 months.
We thus expect that J0840 undergoes stable pulsations of relatively high amplitude, which may facilitate mode identification.
Hence, J0840 offers an excellent laboratory to probe core crystallization via asteroseismology.
J0840 is likely the remnant of a star with a ZAMS mass of $\gtrapprox 7\,M_\odot$ and is thus expected to have a significantly crystallized ONe or ONeMg core.

\citet{corsico2004} suggest that it should be possible to determine the core composition of DAVs based on their pulsation spectrum.
They examine the adiabatic pulsational properties of $1.05\,M_\odot$ WD stars with CO and ONe cores and
find that there are marked differences in the period spacing distributions depending on the core composition.
The ONe-core models displayed significant non-uniformities in the forward period spacing and were also characterized by a larger mean period spacing than CO-models of the same temperature.
The kinetic energy spectra of their ONe-core and CO-core models show significant differences as well.

With three or fewer periods of pulsation detected for these newly discovered DAVs we are unable to conduct an in-depth asteroseismological analyis on these objects.
With future observations and the identification of many more normal modes in the pulsation spectra of these DAVs their total mass and hydrogen envelope mass can be measured.
Further populating the high mass end of the ZZ Ceti instability strip is paramount to the study of crystallization physics 
as the most massive DAVs are expected to have $\sim90$\% crystallized cores.
J0840 is a most interesting DAV as it offers an unprecedented opportunity to constrain the evolution of intermediate-mass stars and the internal structure of a $1.16\,M_\odot$ white dwarf.

\section*{Acknowledgements}
We gratefully acknowledge the support of the NSF and NASA under grants AST-1312678, AST-1312983, and NNX14AF65G.
This paper includes data taken at the McDonald Observatory of The University of Texas at Austin as well as the
Apache Point Observatory 3.5 m telescope, which is owned and operated by the Astrophysical Research Consortium.
This paper is partly based on observations obtained at the Gemini Observatory (acquired through the Gemini Science Archive and processed using the Gemini IRAF package), which is operated by the Association of Universities for Research in Astronomy, Inc., under a cooperative agreement with the NSF on behalf of the 
Gemini partnership: the National Science Foundation (United States), the National Research Council (Canada), CONICYT (Chile), Ministerio de Ciencia, Tecnolog\'{i}a e Innovaci\'{o}n Productiva (Argentina), 
and Minist\'{e}rio da Ci\^{e}ncia, Tecnologia e Inova\c{c}\~{a}o (Brazil).

\bsp	
\label{lastpage}
\end{document}